\documentclass[twocolumn,aps,showpacs]{revtex4}
\usepackage{amssymb}
\usepackage{epsf}

\begin{document}

\title{Asymmetry gap in the electronic band structure of bilayer graphene}
\author{Edward McCann}
\affiliation{Department of Physics, Lancaster University,
Lancaster, LA1 4YB, United Kingdom}

\begin{abstract}
A tight binding model is used to calculate the band structure of
bilayer graphene in the presence of a potential difference between
the layers that opens a gap $\Delta$ between the conduction and
valence bands. In particular, a self consistent Hartree
approximation is used to describe imperfect screening of an
external gate, employed primarily to control the density $n$ of
electrons on the bilayer, resulting in a potential difference
between the layers and a density dependent gap $\Delta (n)$. We
discuss the influence of a finite asymmetry gap $\Delta (0)$ at
zero excess density, caused by the screening of an additional
transverse electric field, on observations of the quantum Hall
effect.
\end{abstract}


\maketitle

Recently there has been a huge amount of theoretical interest in
the low energy electronic properties of ultrathin graphite films,
including graphene monolayers \cite{d+m84,zheng02} and bilayers
\cite{mccann06,nil06}. This activity followed the successful
fabrication of ultrathin graphite films \cite{novo04} and
subsequent measurements of an unusual sequencing of quantum Hall
effect plateaus \cite{novo05,novo06}. While the Hamiltonian for
low energy electrons in a monolayer describes Dirac-like chiral
quasiparticles with Berry phase $\pi$ \cite{d+m84,zheng02}, the
low energy Hamiltonian for a bilayer describes chiral
quasiparticles with a parabolic dispersion and Berry phase $2\pi$
\cite{mccann06}.

\begin{figure}[t]
\centerline{\epsfxsize=\hsize \epsffile{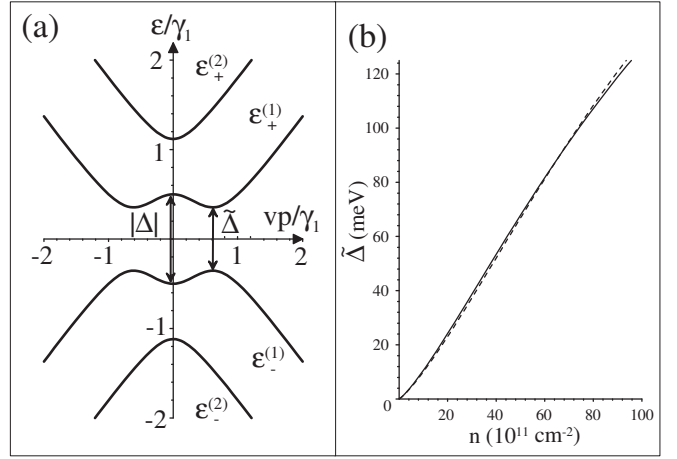}} \caption{(a)
schematic of the electronic bands near the $K$ point in the
presence of finite layer asymmetry $\Delta$ (for illustrative
purposes a very large asymmetry $\Delta = \gamma_1$ is used), (b)
numerically calculated dependence $\tilde{\Delta} (n)$ (solid
line) compared with analytic expression (dashed line) using
Eq.~(\ref{approx1}). For these values of density $\tilde{\Delta}
\approx \Delta$ \cite{truegap}.} \label{fig:1}
\end{figure}

Theoretical \cite{mccann06} and experimental \cite{ohta06} studies
of bilayer graphene have shown that asymmetry $\Delta =
\epsilon_{2} - \epsilon_{1}$ between on-site energies
$\epsilon_{1}$, $\epsilon_{2}$ in the layers leads to a gap
between the otherwise degenerate conduction and valence bands as
shown schematically in Fig.~\ref{fig:1}(a). Moreover, in contrast
to graphene monolayers, there is an experimental possibility of
controlling the magnitude of the gap $\Delta$ in the spectrum of
bilayer graphene through the use of an external gate that is
employed primarily to control the density of electrons $n$ on the
bilayer system \cite{novo04,novo05,novo06,zhang06}. Such a
dependence $\Delta (n)$ owes its existence to the Coulomb
interaction between electrons, but the density distribution itself
is dependent on the value of $\Delta$ via the band structure. Here
we use a self-consistent Hartree approximation to determine the
electronic distribution on a bilayer in the presence of an
external gate. The numerically calculated $\Delta (n)$ is shown in
Fig.~\ref{fig:1}(b) (solid line) for typical experimental
parameters in the case that the gap $\Delta = 0$ at zero excess
density \cite{truegap}. It shows that the addition of density $n
\sim 10^{12}$cm$^{-2}$ yields a gap $\Delta \sim 10$meV. The
numerical curve is compared with the following analytic
approximation (dashed line) valid for $\Delta \ll \gamma_1$:
\begin{eqnarray}
\Delta \approx \frac{e^2 L^2 n}{2 C_b} \left[ 1 + \frac{\Lambda
\hbar^2 v^2 \pi |n|}{\gamma_1^2} - \Lambda \ln \left( \frac{\hbar
v \sqrt{\pi |n|}}{\gamma_1} \right) \right]^{-1} \!\!\!\!\!\! , \,
\label{approx1}
\end{eqnarray}
where $C_b$ is the capacitance of a bilayer of area $L^2$, $v$ is
the in-plane velocity of the bilayer and $\gamma_1$ is the
inter-plane coupling. The dimensionless parameter $\Lambda = e^2
L^2 \gamma_1 / (2\pi\hbar^2 v^2C_b)$ describes the effectiveness
of the screening of the bilayer. The limit $\Lambda \rightarrow 0$
describes poor screening when the density on each layer is equal
to $n/2$ whereas for $\Lambda \rightarrow \infty$ there is
excellent screening, the density lies solely on the layer closest
to the external gate and $\Delta = 0$. For typical experimental
parameters \cite{dressel02} we estimate that $\Lambda \sim 1$.

We consider a graphene bilayer, with interlayer separation $c_0$,
located a distance $d$ from a gate. An external gate voltage $V_g
= e n d / \varepsilon_r^{\prime} \varepsilon_0$ induces a total
excess density $n=n_{1}+n_{2}$ on the bilayer where $n_{1}$
($n_{2}$) is the excess density on the layer closest to (furthest
from) the gate. Here $\varepsilon_{0}$ is the permittivity of free
space, $\varepsilon_{r}^{\prime}$ is the dielectric constant and
$e$ is the electronic charge. Imperfect screening of the effective
charge density $en$ from the gate leads to an excess density
$n_{2}$ on the layer furthest from the gate, with a corresponding
change in potential energy $\Delta
U_{2}=e^{2}n_{2}c_{0}/\varepsilon_{r}\varepsilon _{0}$ that
determines the layer asymmetry
\begin{eqnarray}
\Delta \left( n\right) = \epsilon_{2}-\epsilon_{1} \equiv
\Delta_{0} + e^{2}n_{2}L^2 /C_b \, , \label{deltan}
\end{eqnarray}
where $C_b = \varepsilon_{r} \varepsilon_{0} L^2 /c_0$ and
$\varepsilon_{r}$ is the bilayer dielectric constant. We introduce
the bare asymmetry parameter $\Delta_{0}$ due to an additional
transverse electric field producing finite asymmetry $\Delta (0)$
at zero density. The following analysis includes (a) the tight
binding model of bilayer graphene including the asymmetry
$\Delta$, (b) a self-consistent determination of the gap
numerically and, for $\epsilon_F , \gamma_1 \gg \Delta$,
analytically, and (c) a description of the influence of $\Delta_0$
on the cyclotron mass and the sequencing of quantum Hall effect
plateaus at low density.

\textbf{(a)} The bilayer is modelled as two coupled hexagonal
lattices with inequivalent sites $A1,B1$ and $A2,B2$ in the first
and second graphene sheets, respectively, arranged according to
Bernal ($A2$-$B1$) stacking. This lattice has a degeneracy point
at each of two inequivalent corners, $K$ and $\tilde{K}$, of the
hexagonal Brillouin zone which determine the centers of two
valleys of a gapless spectrum. At the degeneracy point, electron
states on inequivalent ($A1$/$B1$ or $A2$/$B2$) sublattices in a
single layer are decoupled, whereas interlayer coupling
$\gamma_{A2B1} \equiv \gamma_{1}$ forms dimers from pairs of
$A2$-$B1$ orbitals in a bilayer, thus leading to the formation of
high energy bands \cite{mccann06,trig}. The tight-binding
Hamiltonian operates in the space of wave functions $\Psi = (
\psi_{A1} , \psi_{B2} , \psi_{A2} , \psi_{B1} )$ in the valley $K$
and of $\Psi = ( \psi_{B2} , \psi_{A1} , \psi_{B1} , \psi_{A2} )$
in the valley $\tilde{K}$ \cite{mccann06}:
\begin{eqnarray}
\mathcal{H}=\xi \left(
\begin{array}{cccc}
-\frac{1}{2}{\Delta} & 0 & 0 & v{\pi }^{\dag } \\
0 & \frac{1}{2}{\Delta} & v{\pi } & 0 \\
0 & v{\pi }^{\dag } & \frac{1}{2}{\Delta} & \xi \gamma _{1} \\
v{\pi } & 0 & \xi \gamma _{1} & -\frac{1}{2}{\Delta}
\end{array}
\right) ,
\end{eqnarray}
where ${\pi }={p}_{x}+i{p}_{y}$ and $\xi = +1 (-1)$ labels valley
$K$ ($\tilde{K}$). The Hamiltonian takes into account asymmetry
$\Delta = \epsilon_{2} - \epsilon_{1}$ between on-site energies in
the two layers, $\epsilon_{2} = {\textstyle\frac{1}{2}} \Delta$,
$\epsilon_{1} = -{\textstyle\frac{1}{2}} \Delta$. Within each
plane, nearest neighbor coupling as parameterised by $\gamma_0$
leads to in-plane velocity $v=\left( \sqrt{3}/2\right)
a\gamma_{0}/\hbar$ where $a$ is the lattice constant.

At zero magnetic field, the Hamiltonian $\mathcal{H}$ describes
four valley-degenerate bands, $\epsilon_{\pm }^{(\alpha )}
(\mathbf{p})$, $\alpha = 1,2$, with
\begin{eqnarray*}
\epsilon^{(\alpha )2} = \frac{\gamma_{1}^{2}}{2} +
\frac{{\Delta}^{2}}{4} + v^{2} p^{2} +\left( -1\right)^{\alpha }
\! \sqrt{\frac{ \gamma_{1}^{4}}{4} + v^{2}p^{2}\left(
\gamma_{1}^{2} + {\Delta}^{2} \right) },
\end{eqnarray*}%
where $p$ is the magnitude of the momentum near the $K$ point,
Fig.~\ref{fig:1}(a). The energies of the bands exactly at the $K$
point are $|\epsilon_{\pm}^{(2)} (0)| = \sqrt{\gamma_1^2 +
\Delta^2/4}$ and $|\epsilon_{\pm}^{(1)} (0)| = |\Delta| /2$. Thus
$\epsilon_{\pm}^{(2)}$ describes higher-energy ($A2$-$B1$ dimer)
bands $|\epsilon_{\pm}^{(2)}| \geq \gamma_1$ whereas
$\epsilon_{\pm}^{(1)}$ are low energy bands split by the layer
asymmetry $\Delta$ at the $K$ point \cite{truegap}.

The electronic densities $n_{1}$ and $n_{2}$ on the individual
layers are given by an integral with respect to momentum over the
circular Fermi surface $n_{1(2)} = (2 /\pi \hbar^2) \int p \, dp
\left( | \psi_{A1(2)} ( p ) |^2 + | \psi_{B1(2)} ( p ) |^2
\right)$ where we have included a factor of four to take into
account spin and valley degeneracy. On determining the
wavefunction amplitudes on the four atomic sites we find
\begin{eqnarray}
n_{1(2)} &=&  \int d p  \, p \, g_{\mp} ( \epsilon , p ) \, ,
\label{n12} \\
g_{\mp} ( \epsilon , p ) \!&=&\! \frac{ \epsilon \mp \Delta/2
}{\pi \hbar^2 \epsilon} \left[ \frac{ \left( \epsilon^2 - \Delta^2
/4 \right)^2 \mp 2 v^2 p^2 \epsilon \Delta - v^4 p^4 } { \left(
\epsilon^2 - \Delta^2 /4 \right)^2 + v^2 p^2 \Delta^2 - v^4 p^4 }
\right] , \nonumber
\end{eqnarray}
where the minus (plus) sign is for the first (second) layer. The
limits of integration are allowed values of momentum that,
depending on the band in question and the value of the Fermi
energy $\epsilon_F$, are $p=0$ or $p_{\pm}$ where $v^2 p_{\pm}^2 =
\epsilon_F^2 + {\textstyle\frac{1}{4}}\Delta^2 \pm [\epsilon_F^2 (
\gamma_1^2 + \Delta^2 ) - {\textstyle\frac{1}{4}} \gamma_1^2
\Delta^2]^{1/2}$.

\begin{figure}[t]
\centerline{\epsfxsize=\hsize \epsffile{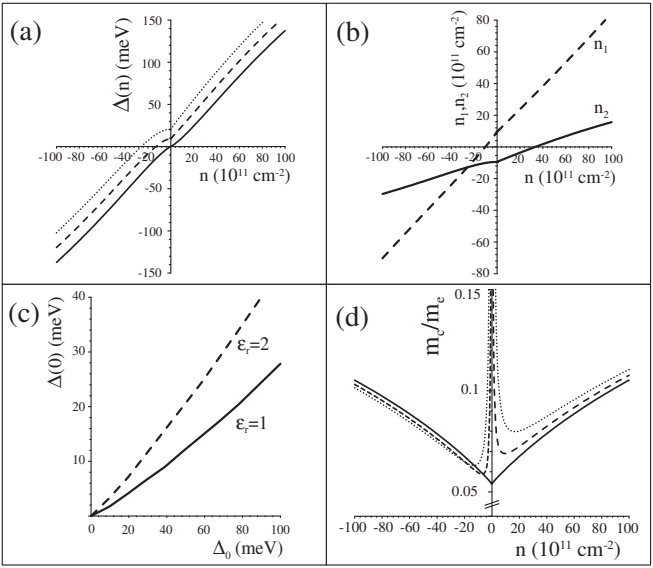}}
\caption{\label{fig:2} Numerical evaluation of (a) the bilayer
asymmetry $\Delta (n)$ for different values of the bare asymmetry
$\Delta_0 = 0$ (solid line), $\Delta_0 = 0.1\gamma_1 = 39$meV
(dashed line) and $\Delta_0 = 0.2\gamma_1 = 78$meV (dotted line),
using typical parameter values \cite{dressel02}, (b) layer
densities $n_2$ (solid) and $n_1$ (dashed) as a function of $n$
for $\Delta_0 = 0.2\gamma_1 = 78$meV, (c) $\Delta (0)$ as a
function of $\Delta_0$ for $\varepsilon_{r} = 1$ (solid line) and
$\varepsilon_{r} = 2$ (dashed), and (d) the cyclotron mass $m_c$
in units of the bare electronic mass $m_e$, for different values
of $\Delta_0$ as in (a).}
\end{figure}

\textbf{(b)} The self-consistent calculation begins with zero
external gate voltage, assuming the Fermi energy to be located at
$\epsilon =0$. We take into account the contribution of the two
valence bands to the initial densities on each layer, $n_{1}^{(0)}
= - n_{2}^{(0)}$, and evaluate the gap $\Delta ( 0 )$ using
Eqs.~(\ref{deltan},\ref{n12}). Then, we proceed to the case where
an external gate voltage produces excess densities $n_{1}$,
$n_{2}$, $n = n_{1}+n_{2} \neq 0$, and a gap $\Delta ( n )$.
Figure~\ref{fig:2}(a) shows $\Delta (n)$ for different values of
the bare asymmetry $\Delta_0$ \cite{truegap} and
Fig.~\ref{fig:2}(b) shows the variation of the individual layer
densities $n_1$ and $n_2$ for finite $\Delta_0$. The main effect
of finite $\Delta_0$ (dashed and dotted lines in
Fig.~\ref{fig:2}(a)) is to shift the plot $\Delta (n)$ with
respect to the $\Delta_0 = 0$ case (solid line), but, as a result
of screening, the asymmetry at zero density $\Delta (0)$ is
smaller than the bare asymmetry $\Delta_0$. For example, the
dotted curve for $\Delta_0 = 0.2 \gamma_1 = 78$meV shows $\Delta
(0) \approx 21$meV. Figure~\ref{fig:2}(c) shows the variation of
$\Delta (0)$ with bare asymmetry $\Delta_0$ (solid line).

We find an analytic expression for $\Delta (n)$ in the limit
$\gamma_{1} , \epsilon _{F} \gg |\Delta |$ by evaluating the
density on each layer Eq.~(\ref{n12}) using an expansion in
$\Delta / \gamma_{1}$. For simplicity, we consider $|\Delta
|/2<|\epsilon_{F}|<\gamma _{1}$ so that either the band just above
or just below zero energy is partially occupied. On integrating
Eq.~(\ref{n12}) from zero momentum to $p_F = p_{+}$, we find the
densities in the partially occupied bands. We also take into
account the redistribution of density within the valence bands for
$\epsilon _{F} \neq 0$ as compared to the $\epsilon _{F}=0 $\
case, as given by $\pm \gamma _{1}\Delta /\left( 4\pi \hbar^{2}
v^{2}\right) \ln \left( 4\gamma _{1}/|\Delta |\right)$, so that
\begin{eqnarray*}
n_{1\left( 2\right) }\approx \frac{\mathrm{sgn}\left( \epsilon
_{F}\right) p_{F}^{2}}{2\pi \hbar ^{2}} \pm \frac{\gamma _{1}
\Delta }{2\pi \hbar ^{2}v^{2}} \left[ \frac{\epsilon_0}{\gamma_1}
+ \frac{\epsilon_0^2}{\gamma_1^2} - {\textstyle\frac{1}{2}} \ln
\left( \frac{\epsilon_0}{\gamma_1}\right) \right] ,
\end{eqnarray*}
where $\epsilon_0 = (\gamma_1 /2 ) [\sqrt{1+4v^{2}p_{F}^{2}/\gamma
_{1}^{2}}-1 ]$ and $n=n_{1}+n_{2}\approx \mathrm{sgn}\left(
\epsilon _{F}\right) p_{F}^{2}/\left( \pi \hbar ^{2}\right)$. If
$|\Delta |\ll \{|\epsilon _{F}|,\gamma _{1}\}$ so that
$v^{2}p_{F}^{2}\approx \epsilon_{F}^{2} + \gamma_{1}|\epsilon
_{F}|$, then $\epsilon_0$ is approximately independent of
$\Delta$, and we may use Eq.~(\ref{deltan}) to find the gap,
\begin{eqnarray}
\Delta  &\approx&  \frac{\Delta_0 + e^2 L^2 n /(2C_b) } {1 +
\Lambda \left[ ( \epsilon_0 / \gamma_1 ) + ( \epsilon_0 / \gamma_1
)^2 - {\textstyle\frac{1}{2}} \ln \left( \epsilon_0 / \gamma_1
\right) \right]} . \label{approx2}
\end{eqnarray}
This is valid for intermediate densities $|\Delta | , |\Delta_0|
\ll |\epsilon_F| < \gamma_1$ where $\epsilon_F \approx \pm
\epsilon_0$ [at very low density it incorrectly predicts $\Delta
(0) = 0$ for $\Delta_0 \neq 0$]. In the regime $4 \pi \hbar^2 v^2
|n| \ll \gamma_1^2$ and $\Delta_0 = 0$ it gives
Eq.~(\ref{approx1}).

%
%
%
%

There is uncertainty in the value of the capacitance $C_b$ within
this model because we assume that the excess charge is uniformly
distributed within infinitesimally thin 2d layers. Also, we used
$\varepsilon_r =1$ for the dielectric constant of the bilayer
which agrees with the prediction for other low dimensional
structures \cite{dcsmall}, but is smaller than the value for bulk
graphite $\varepsilon_r \approx 2.4$ \cite{dcbulk}. Furthermore,
we assume that the model parameters, including layer separation
and interlayer coupling $\gamma_1$, do not vary as a function of
applied potential whereas a recent experiment \cite{ohta06} found
that $\gamma_1$ changed by about $3$\% as $\Delta$ changed by
$100$meV. The effectiveness of screening $\Lambda = e^2 L^2
\gamma_1 / (2\pi\hbar^2 v^2C_b)$ depends on the model parameters
as illustrated in Fig.~\ref{fig:2}(c) which shows $\Delta (0)$ as
a function of $\Delta_0$ for $\varepsilon_{r} = 1$ (solid line)
and $\varepsilon_{r} = 2$ (dashed): the gap $\Delta (0)$ increases
with the dielectric constant.

\textbf{(c)} We use the self consistent analysis to evaluate the
semiclassical expression for cyclotron mass $m_{c}$ that may be
measured in an experimental observation at finite magnetic field
of, say, Shubnikov de Haas oscillations
\cite{novo04,novo05,novo06}. For a circular Fermi surface, the
semiclassical expression for cyclotron mass $m_{c}= p / (\partial
\epsilon /\partial p )$ gives
\begin{eqnarray*}
m_{c} = \frac{\epsilon^{(\alpha})}{v^{2}}\left[ 1 + (-1)^{\alpha}
\frac{\gamma _{1}^{2}+\Delta ^{2}}{2\sqrt{v^{2}p^{2}\left[ \gamma
_{1}^{2}+\Delta ^{2}\right] +\gamma _{1}^{4}/4}}\right] ^{-1} .
\end{eqnarray*}
In the limit $\Delta = 0$ for the low energy bands $\alpha = 1$
this gives $m_{c}= ( \gamma_{1}/2v^{2}) \sqrt{1+ 4\pi \hbar
^{2}v^{2}|n|/\gamma _{1}^{2}}$ \cite{mccann06}.
Figure~\ref{fig:2}(d) shows the cyclotron mass $m_{c}$ as a
function of total density $n$. For $\Delta_0 = 0$ (solid line) the
mass is symmetric and finite $m_c = \gamma _{1}/2v^{2}$ at zero
density. The other curves, for $\Delta_0 > 0$, show that the
cyclotron mass for positive and negative densities is asymmetric
with divergent behavior at low density resulting from the
\textquotedblleft mexican hat\textquotedblright\ structure of the
low energy bands \cite{trig}.

Another manifestation of finite $\Delta_0$ is in the sequencing of
quantum Hall effect (QHE) plateaus at low density. The Landau
level (LL) spectrum of bilayer graphene has been discussed in
Ref.~\cite{mccann06}: here we briefly mention the features
relevant for finite $\Delta_0$. The low energy states of electrons
in bands $\epsilon_{\pm}^{(1)}$ are conveniently described by an
effective two component Hamiltonian that operates in the space of
wave functions $\Psi = ( \psi_{A1} , \psi_{B2} )$ in the valley
$K$ and of $\Psi = ( \psi_{B2} , \psi_{A1} )$ in the valley
$\tilde{K}$ \cite{mccann06}:
\begin{eqnarray*}
{\hat{H}} &=& - \, \frac{1}{2m}\left(
\begin{array}{cc}
0 & \left( {\pi }^{\dag }\right) ^{2} \\
{\pi ^{2}} & 0
\end{array}
\right) \\
&& \, - \, \xi \Delta \left[ {\textstyle\frac{1}{2}}\left(
\begin{array}{cc}
1 & 0 \\
0 & -1%
\end{array}%
\right) -\frac{v^{2}}{\gamma _{1}^{2}}\left(
\begin{array}{cc}
{\pi }^{\dag }{\pi } & 0 \\
0 & -{\pi \pi }^{\dag }
\end{array}
\right) \right] \\
&& \, - \, \frac{\alpha}{2m} \left(
\begin{array}{cc}
p^2 & 0 \\
0 & p^2
\end{array}
\right) + \frac{\beta \hbar e B}{2m} \left(
\begin{array}{cc}
1 & 0 \\
0 & -1
\end{array}
\right) ,
\end{eqnarray*}
where ${\pi }={p}_{x}+i{p}_{y}$ and $B$ is the magnetic field. For
completeness we include the final two terms, with dimensionless
parameters $|\alpha |, |\beta | \ll 1$, describing weak asymmetry
between the conduction and valence bands. Neglected in
Ref.~\cite{mccann06}, they arise due to additional couplings such
as next-to-nearest neighbor in-plane coupling between $A1$-$A1$
and $B2$-$B2$ orbitals or interlayer coupling between $A1$-$A2$
and $B1$-$B2$ orbitals. The Landau level (LL) spectrum may be
found using the Landau gauge $\mathbf{A}=\left( 0,Bx\right)$, in
which operators ${\pi }^{\dag }$ and $\pi$ coincide with raising
and lowering operators \cite{p+r87} in the basis of Landau
functions $e^{iky}\phi _{N}(x)$, such that $\pi ^{\dag }\phi
_{N}=i(\hbar /\lambda _{B})\sqrt{2(N+1)}\phi _{N+1}$, $\pi \,\phi
_{N}=-i(\hbar /\lambda _{B})\sqrt{2N}\phi _{N-1}$, and $\pi \,\phi
_{0}=0$, where $\lambda _{B}=\sqrt{\hbar /(eB)} $. The spectrum
has almost equidistant energy levels for $N \geq 2$ with spacing
$\hbar \omega_c$, $\omega_{c}=eB/m$, which are weakly split in
valleys $K$ and $\tilde{K}$ ($\xi =\pm 1$) \cite{mccann06},
\begin{eqnarray}
\!\!\!\!\!\! \epsilon_{N \geq 2}^{\pm } &\approx& \frac{\xi \Delta
\hbar \omega_c}{2 \gamma_1} - \alpha \hbar \omega_c (N -
{\textstyle\frac{1}{2}}) \pm \hbar \omega _{c}\sqrt{N(N-1)}
 , \label{LLhigh}
\end{eqnarray}
where $\epsilon_{N}^{+}$ and $\epsilon_{N}^{-}$ are assigned to
electron and hole states, respectively, and we assume $|\Delta | /
\gamma_1 \ll 1$, $|\Delta | / \hbar \omega_c \alt 1$.

\begin{figure}[t]
\centerline{\epsfxsize=1.0\hsize\epsffile{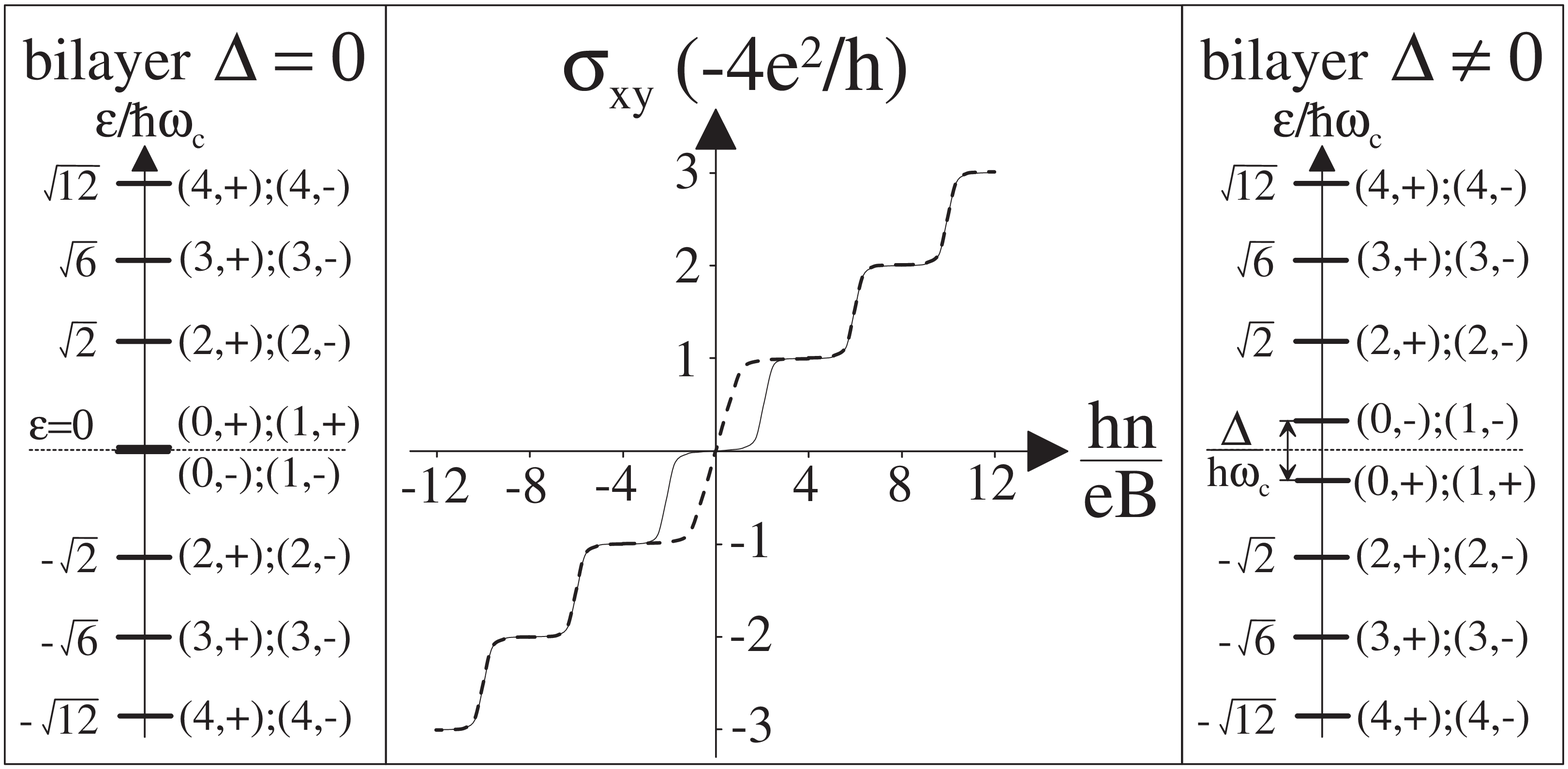}}
\caption{Landau levels for zero (left) and finite asymmetry
$\Delta$ (right). Brackets $(N,\protect\xi )$ indicate LL number
$N$ and valley index $\protect\xi =\pm 1 $. In the center the
predicted Hall conductivity $\protect\sigma _{xy}$ as a function
of carrier density for bare asymmetry $\Delta_0 = 0$ (dashed line)
is compared to that for $\Delta_0 \neq 0$ (solid line).}
\label{fig:3}
\end{figure}

The LL spectrum in each valley also contains two levels identified
using the fact that ${\pi ^{2}} \phi_{1}={\pi ^{2}}\phi _{0}=0$,
\begin{eqnarray*}
\left\{
\begin{array}{ll}
\epsilon_{0} = -{\frac{1}{2}}\xi \Delta - {\frac{1}{2}} ( \alpha -
\beta) \hbar \omega_c \,; & \quad \Phi_{0 \xi }\equiv
(\phi _{0},0); \\
\epsilon_{1} = - {\frac{1}{2}}\xi \Delta + \frac{\xi \Delta \hbar
\omega_c}{\gamma_1} - {\frac{1}{2}} ( 3\alpha - \beta) \hbar
\omega_c \,; & \quad \Phi _{1\xi }\equiv (\phi_{1},0).
\end{array}
\right.
\end{eqnarray*}
According to different definitions of two-component $\Phi $ in two
valleys, $N=0,1$ LL states in the valley $K$ are formed by
orbitals predominantly on the $A$ sites from the bottom layer,
whereas the corresponding states in the valley $\tilde{K}$ are
located on $\tilde{B}$ sites from the top layer, as reflected by
the splitting $\Delta$ between the lowest LL in the two valleys.
The spectrum is shown on the right side of Fig.~\ref{fig:3}
assuming $|\Delta | / \gamma_1 \ll 1$, $|\Delta | / \hbar \omega_c
\alt 1$. In a symmetric bilayer ($\Delta = 0$) levels
$\epsilon_{0}$\ and $\epsilon _{1}$\ are degenerate and have the
same energy in valleys $K$ and $\tilde{K}$, thus forming an 8-fold
degenerate LL at $\epsilon =0$ (here, spin is taken into account)
as shown on the left side of Fig.~\ref{fig:3}.

%
%
%
%

The LL spectrum is reflected in the QHE Hall conductivity
dependence on carrier density, $\sigma _{xy}(n)$, although, in
general, finite temperature and LL broadening tend to mask small
LL splitting. When the asymmetry at zero density $\Delta (0)$,
Fig.~\ref{fig:2}(c), is large enough that the percolating states
\cite{p+r87} from the lowest levels in the two valleys can be
resolved (the right side of Fig.~\ref{fig:3}), then these levels,
embedded into the ladder of 4-fold degenerate LL's with $n\geq 2$,
Eq. (\ref{LLhigh}), result in the form of $\sigma_{xy}(n)$ as
sketched in the center of Fig.~\ref{fig:3} (solid line) which
exhibits plateaus at all integer values of $4e^{2}/h$ including a
plateau at zero density. The temperature dependence of the zero
density plateau will differ from that of the other plateaus
because of the different activation energies, related to $\Delta
(0)$ and $\hbar \omega_c$, respectively. When it is not possible
to resolve the asymmetry splitting of the group of 8 states at
$\left\vert \epsilon \right\vert =0$ (for $\Delta (0) \approx
\Delta_0 \approx 0$) then $\sigma_{xy}(n)$ (dashed line in
Fig.~\ref{fig:3}) has plateaus at integer values of $4e^{2}/h$ and
a double $8e^{2}/h$ step across $n=0$ \cite{mccann06,novo06}.

The use of doping \cite{ohta06} recently achieved gaps of up to
about $200$meV. Alternatively, a potential difference of $30$V
applied to a back gate placed $300$nm from the bilayer (combined
with $-3$V applied to a top gate $30$nm away) translates into
$30$mV across the bilayer, $\Delta_0 = 30$meV, and $\Delta (0)
\approx 7$meV. This compares with $\hbar \omega_c \approx 22$meV
for field $B = 10$T \cite{dressel02} and level broadening due to
scattering of approximately $3$meV in recent experiments
\cite{zhang06}.

This paper has considered the influence of a single gate placed
above an ultra thin graphite film as employed in recent
experiments \cite{novo04,novo05,novo06,zhang06} to control the
density of electrons $n$ on the film. For the typical experimental
range of densities shown in Fig.~\ref{fig:1}(b) the dependence of
the asymmetry gap $\Delta (n)$ is roughly linear with $n$ with the
addition of density $n \sim 10^{12}$cm$^{-2}$ yielding a gap
$\Delta \sim 10$meV. Finite bilayer asymmetry gap $\Delta (0)$ at
zero excess density, due to an additional transverse electric
field, could influence observations of the integer quantum Hall
effect by introducing a plateau at zero density in the Hall
conductivity. This plateau is accompanied by a dip in the diagonal
conductivity, the temperature dependence of which can be used to
extract the value of the asymmetry gap $\Delta (0)$. This is
distinguishable from the behavior of the higher plateaus with
activation energies determined by the cyclotron frequency.

\acknowledgments The author thanks V.~I.~Fal'ko, A.~K.~Geim and
L.~M.~K.~Vandersypen for useful discussions and EPSRC for
financial support.



\begin{thebibliography}{99}
\bibitem{truegap}
The true value of the gap between the conduction and valence band,
occuring at $p \neq 0$, is $\tilde{\Delta} = |\Delta |\gamma_1 /
(\gamma_1^2 + \Delta^2)^{1/2} $. For huge asymmetry $|\Delta | \gg
\gamma_1$, the gap saturates at $\tilde{\Delta} \approx \gamma_1$
although for modest asymmetry $|\Delta | \ll \gamma_1$, as
considered throughout this paper, we have $\tilde{\Delta} \approx
|\Delta |$.
\bibitem{trig} We neglect the role
of additional weak couplings, such as $A1$-$B2$ coupling
$\gamma_{3}$ that results in trigonal warping and is relevant at
low density $n \sim 1 \times 10^{11}$cm$^{-2}$ \cite{mccann06}.
\end{thebibliography}
\end{document}